# Excited states engineering maximizes singlet generation by triplet fusion in conjugated systems.

Alessandra Ronchi,[1] Masashi Mamada,[2] Michel Frigoli,[3*] and Angelo Monguzzi[1*]

[1] Department of Material Science, University of Milano-Bicocca, via R. Cozzi 55, 20125 Milan, Italy

[2] Department of Chemistry, Graduate School of Science, Kyoto University, Kitashirakawa Oiwake-cho, Sakyo-ku, Kyoto 606–8502, Japan

[3] UMR CNRS 8180, UVSQ, Institut Lavoisier de Versailles, Université Paris-Saclay, 45 Avenue des Etats-Unis, 78035 Versailles Cedex, France

E-mail: michel.frigoli@uvsq.fr , angelo.monguzzi@unimib.it

ABSTRACT

Photon upconverters are anti-Stokes emitters capable of generating photons with higher energy than those absorbed. This behavior can be achieved through different mechanisms, which are extensively studied for applications in imaging, anticounterfeiting, information encryption and most importantly, solar energy technologies. Among these mechanisms, photon upconversion based on sensitized triplet-triplet annihilation (sTTA-UC) is particularly attractive because it operates under low-intensity, incoherent light. In sTTA-UC, two optically dark triplet states of a conjugated annihilator fuse upon collision to populate a higher-energy fluorescent singlet, with the triplets initially generated via energy transfer from a lower-energy sensitizer. Here we introduce a general molecular-design strategy to maximize singlet generation through TTA. By selective substitution, we engineered a naphthalene-derived annihilator with an excited-state energy landscape that strongly favors singlet formation, achieving yields up to 0.83. When combined with an appropriate triplet sensitizer, the system delivers stable UV–visible upconversion peaking at 390 nm, with an absolute upconversion yield of 0.19 and an activation excitation intensity threshold lower than 0.1 suns under non-coherent broadband excitation fully compatible with the requirements of solar-powered technologies.

Anti-Stokes emitters are materials that, upon absorption of excitation light, show photoluminescence at energy higher than that of the absorbed photons.[1] Such emitters are commonly referred to as photon upconverters, as they generate photons with energies exceeding those of the incident radiation. This peculiar behavior can be a consequence of several mechanisms, including phonon absorption, non-linear optical properties, sequential and simultaneous multiple absorptions and triplet-triplet annihilation (TTA). All these mechanisms are extensively studied not only for the general interest in the fundamental physics involved, but also for their potential application in imaging,[2-6] anticounterfeiting,[7-10] information encryption[11-13] and importantly, solar energy-based technologies.[1,14-16] The energy loss affecting photovoltaic and photocatalytic devices is indeed most largely due to the light harvesting semiconductor's inability to absorb the low energy tail of the solar emission.[17,18] This still strongly limits the absolute amount of electrical power or catalytic efficacy that can be achieved, with detrimental consequences on the broad diffusion of solar devices for green energy production. To establish whether the implementation of photon upconversion could beneficially impact real life technologies, it is essential to quantify the efficiency increment that an upconverter can provide to a solar device. The photon conversion quantum yield $\phi_{uc}$ is therefore a crucial parameter to assess the applicability of any upconverter.

Unlike two-photon absorption, second-harmonic generation, and excited-state absorption in lanthanide-based materials,[19-21] photon upconversion based on sensitized TTA (sTTA-UC) can be used to blue-shift non-coherent and low-intensity radiation, so in principle it can be used to manage solar photons.[22-26] As shown in Fig. 1a, in sTTA-UC the upconverted luminescence originates from the fusion upon collision of two optically dark triplets of an annihilator/emitter, followed by the population of a high-energy fluorescent singlet state. The emitter triplets are populated by Dexter energy transfer from the triplets of a low-energy light-harvesting moiety, i.e. the sensitizer.[27,28] The process quantum efficiency $\phi_{uc}$ can be defined as the number of upconverted photons emitted out of the number of excitation photons absorbed by the sensitizer, so

$$\phi_{uc} = \#ph_{em}/\#ph_{abs} \ , \qquad \text{Eq. 1}$$

where $\#ph_{em} = \int_{-\infty}^{+\infty} PL(\lambda) d\lambda$ is the wavelength-integrated photoluminescence spectrum $PL(\lambda)$ recorded with a photon counter and $\#ph_{abs} = \int_{-\infty}^{+\infty} \bar{\alpha}(\lambda) I_{exc}(\lambda)\, d\lambda$ takes into account the sensitizer's absorptance $\bar{\alpha}(\lambda)$ and the emission spectrum $I_{exc}(\lambda)$ of the excitation source. For sTTA-UC Eq. 1 evolves to

$$\phi_{uc}[n, I_{exc}] = \frac{\prod_i \phi_i}{n} \xrightarrow{sTTA-UC} \frac{\phi_{ISC}\phi_{ET}\phi_{fl}\phi_{T\to S}}{n} \phi_{TTA}(I_{exc}) = \mathrm{f}\frac{\phi_{ISC}\phi_{ET}\phi_{fl}}{n} \phi_{TTA}(I_{exc}) \qquad \text{Eq. 2}$$

where the index $i$ runs over the photophysical steps involved and $n = 2$ is the number of photons absorbed for each photon emitted. The numerator in Eq. 2 is the product of the efficiencies of intersystem crossing on sensitizers ($\phi_{ISC}$), energy transfer ($\phi_{ET}$), emitter fluorescence ($\phi_{fl}$), TTA ($\phi_{TTA}$) and singlet generation upon TTA ($\phi_{T\to S}$). The only power-dependent factor in the equation is $\phi_{TTA}$, which is set by the triplets density and collisional rate.[29,30] In the ideal case where $\prod_i \phi_i$=1, the energy conservation law sets the maximum conversion efficiency achievable $\phi_{uc}^{max}$ to 0.50. In the framework of TTA, the $\phi_{T\to S}$ value is usually constant and reported as the statistical factor f, which indicates the probability to form the lowest excited singlet $S_1$ upon the annihilation of two emitter triplets $T_1$. The f value is usually much lower than unity. For common annihilators, it is typically assessed between 0.2 and 0.5.[31-35] While a lot of effort has been dedicated to develop new triplet sensitizers including semiconductor quantum dots and perovskites,[28,36-41] few works were devoted to the investigation of this aspect,[31,42-45] which is still a critical bottleneck especially towards the development of upconverters working in the UV[32,33,46-50] and NIR[51-53] ranges where the coupling with existing solar-powered technologies could give the most significant efficiency improvement.

We propose here a general strategy to maximize the statistical singlet generation yield through TTA by engineering the electronic states of a conjugated annihilator. By exploiting the effect of selective substitution, a naphthalene-derived annihilator has been designed to have an electronic energy distribution which favors the formation of the bright excited singlets, showing an effective f value as high as 0.83. By combining this annihilator with an appropriate triplet sensitizer, we achieved highly stable upconverted emission in the UV-Vis at 390 nm with $\phi_{uc}$ ~ 0.19, and an excitation intensity threshold for the sTTA-UC process at ~ 0.7 mW cm$^{-2}$ (< 0.1 suns) under non-coherent broadband excitation, making the system fully compatible with the technological demands of real life solar-powered devices.

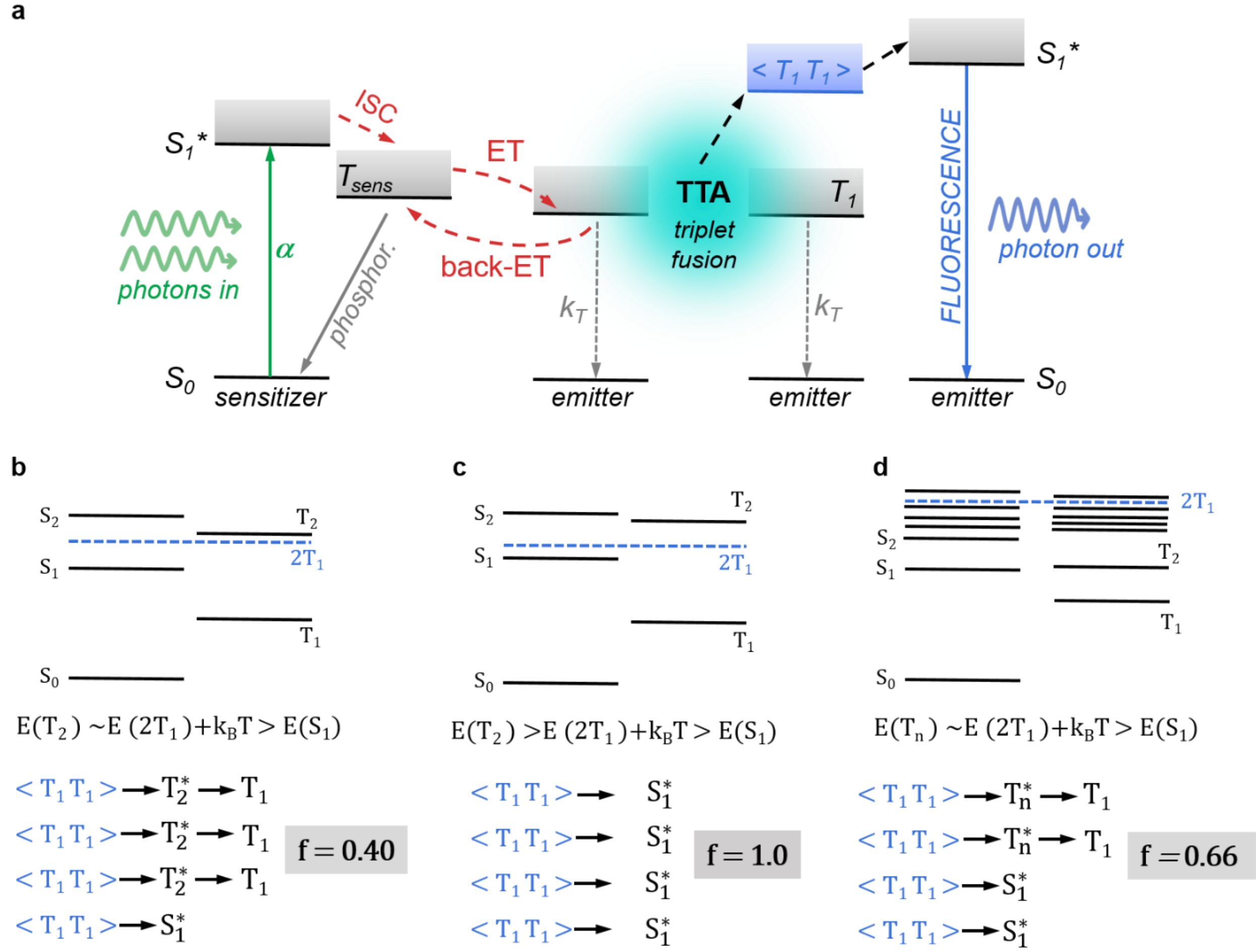

**Fig. 1 | General photophysics of sensitized triplet-triplet annihilation based photon upconversion (sTTA-UC). a)** Following photon absorption, a light harvester/triplet sensitizer molecule with absorption coefficient α is promoted to its excited singlet state ($S_1^*$) that efficiently undergoes intersystem crossing (ISC) to its triplet state $T_1$ ($T_{sens}$). Forward energy transfer (ET) occurs and competes with back-energy transfer (back-ET) to the annihilator/emitter triplet state $T_1$. This can either spontaneously decay or undergo triplet–triplet annihilation (TTA). The triplets collisional complex $<T_1T_1>$ can relax to the emitter $S_1$ state, which then emits high energy light when relaxing to the ground state. Dashed lines indicate radiation-less transitions. **b-d)** Sketch of the TTA process involving four triplet pairs. In case (b) the emitter $T_2$ state is energetically accessible at room temperature, therefore the collision of four triplet pairs (8 triplets in total) destroys five triplets. Only two dead triplets produce an excited singlet state and f = 2/5 = 0.40. In the ideal case (c), the $T_2$ level is energetically inaccessible. The collision of four triplet pairs produces four singlet excited states, with f = 8/8 = 1.00. In case (d), the $<T_1T_1>$ energy is strictly resonant with both high energy singlet and triplet states, and both channels are equally accessible for the $<T_1T_1>$ relaxation. If ISC is neglected, four out of six dead triplets populate two excited singlets, with f = 4/6 = 0.66.

As sketched in Fig. 1b-d, the statistical factor f is mainly set by the distribution and resonance between the annihilator's singlet and triplet states. A low f value is often observed because the $T_2$ level is typically resonant with the energy of the collisional complex $<T_1T_1>$ formed upon TTA. The possible non-radiative relaxation of $T_2$ to $T_1$ poses a loss channel because of the competitive formation of non-emissive triplets, with a maximum theoretical f of 0.4 (Fig. 1b).[54] On the other hand, the ideal annihilator's electronic structure is found when the $<T_1T_1>$ energy is slightly higher than $S_1$, and the $T_2$ is too energetic to be reached (Fig. 1c). In this case the $<T_1T_1>$ complex can only relax to the emissive $S_1$ state, with f = 1. However, TTA annihilators are usually small conjugated molecules that are particularly sensitive to structural changes, so the desired configuration with a broad distribution of well separated energy levels is exceptionally difficult to achieve. Therefore, instead of trying to minimize the energetic resonance between the $<T_1T_1>$ complex and $T_n$ states to reduce energy losses, we propose a different approach by designing an annihilator with high resonance between the $<T_1T_1>$ complex and both a $T_n$ and $S_n$ states. In this case, considering an equal probability to populate singlets and triplets by $<T_1T_1>$ relaxation, four out of six dead triplets can produce singlets, with an expected intrinsic f value as large as 0.66 (Fig. 1d). These considerations inspired the design of a new and optimized annihilator/emitter.

Figure 2a shows the molecular structure of 1,5-bis[(triisopropylsilyl)ethynyl]--3,7-diphenylnaphthalene (TIPS-Ph-Naph) we designed and synthesized as annihilator as well as the structure of the phosphorescent Ir complex employed as light harvester and triplet sensitizer (bis[2-(2-pyridinyl-N)phenyl-C](acetylacetonato)iridium(III) ($Ir(ppy)_2(acac)$) (Supplementary Fig. S1). The TIPS-Ph-Naph has been synthesized by a Sonogashira coupling between 1,5-bis(trifluoromethanesulfonate)-3,7-diphenylnaphthalene and (Triisopropylsilyl)acetylene (Supplementary section 1). Lateral substitutions have been introduced to minimize the luminescence quenching in naphthalene (Naph) and tune its electronic structure so as to exploit its triplet state for TTA, as previously demonstrated with TIPS groups installed at the 1,4- or 1,5-positions of the naphthalene core.[32,33,55,56]

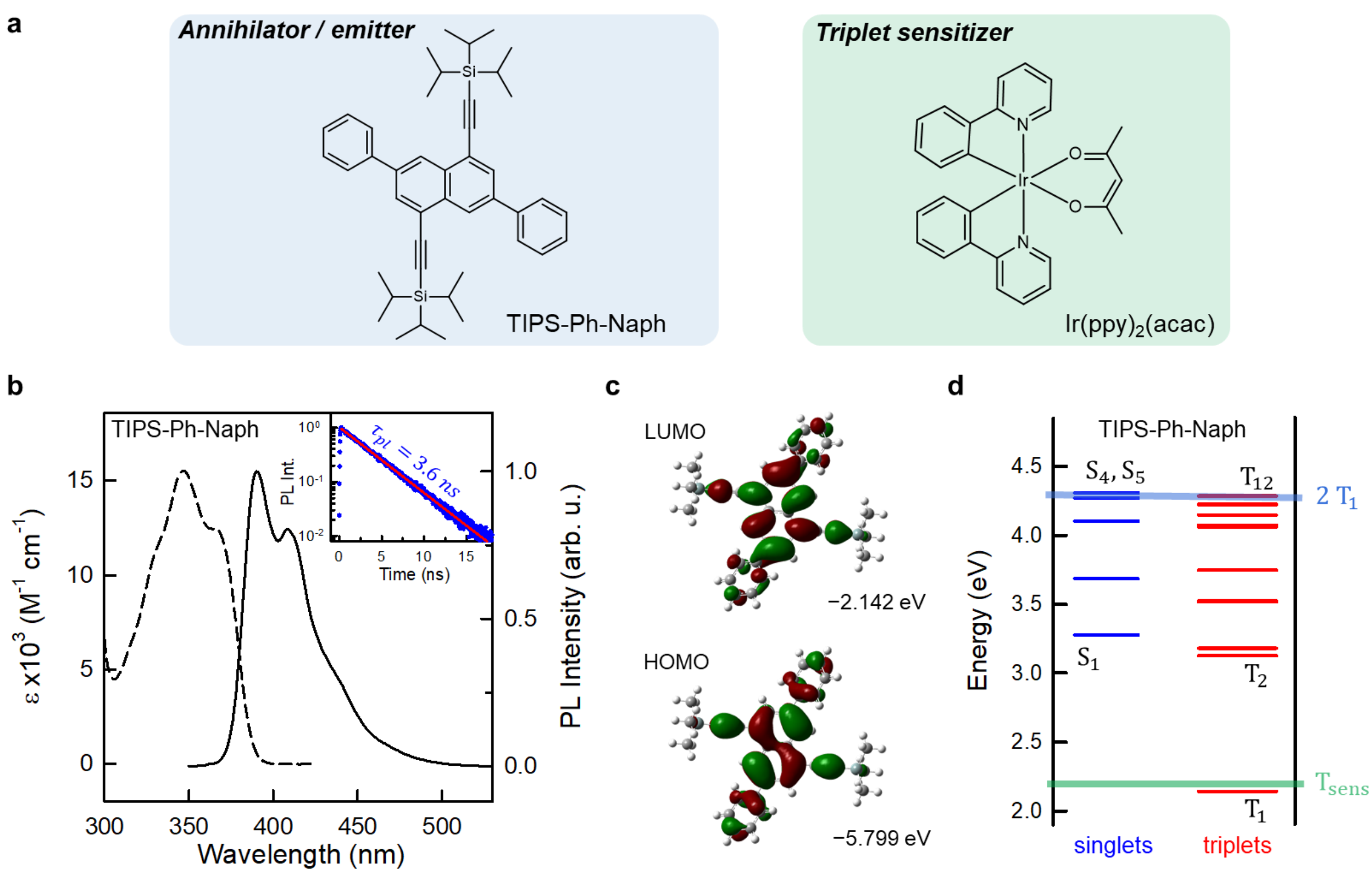

**Fig. 2 | Absorption, photoluminescence and electronic properties of the proposed naphthalene derivative TIPS-Ph-Naph employed as triplet-triplet annihilator**. **a)** Molecular structure of the selected triplet sensitizer $Ir(ppy)_2(acac)$ and of the naphthalene-derivative emitter TIPS-Ph-Naph. **b)** Molar extinction coefficient (dashed line) and photoluminescence (PL, solid line) spectra of TIPS-Ph-Naph in THF solution. The PL spectrum was acquired under cw excitation at 320 nm. The inset reports the PL intensity decay trace at 390 nm recorded under pulsed laser excitation at 340 nm. The solid line is the fit of data with a single exponential decay function with characteristic decay time $\tau_{pl}$ = 3.6 ns. **c)** Shapes and energies of the highest occupied molecular orbital (HOMO) and lowest unoccupied molecular orbital (LUMO) of TIPS-Ph-Naph calculated by DFT at the B3LYP/6-311+G(d,p) level. **d)** TD-DFT calculations of singlet and triplet energies of TIPS-Ph-Naph at the B3LYP/6-311+G(d,p) level. As a reference, we also highlighted the $Ir(ppy)_2(acac)$ triplet energy at 2.14 eV (green stripe) and the energy of the collisional complex $2T_1$ (blue stripe).

Figure 2b shows the absorption and photoluminescence properties of the TIPS-Ph-Naph emitter, in which TIPS groups and phenyl rings are incorporated at the centrosymmetric 1,5- and 3,7-positions of the naphthalene core, respectively. The molecule displays a clearer vibronic structure in THF solution compared to Naph (Supplementary Fig. S2), a $\phi_{pl}$ of 0.59±0.04 and characteristic fluorescence lifetime $\tau_{pl}$ = 3.6 ns (Fig. 2b, inset). Notably, lowering the temperature to 77 K does not affect the emission lifetime, but it better resolves the molecular vibronic structure due to partial suppression of thermal broadening (Supplementary Fig. S5). We ascribe this behavior to a temperature independent ISC from $S_1$ to $T_1$, which limits the TIPS-Ph-Naph fluorescence quantum yield.[57] TIPS-Ph-Naph shows a $T_1$ energy of 2.12 eV (Supplementary Fig. S5), in excellent agreement with the calculated value (2.14 eV, Fig. 2d,

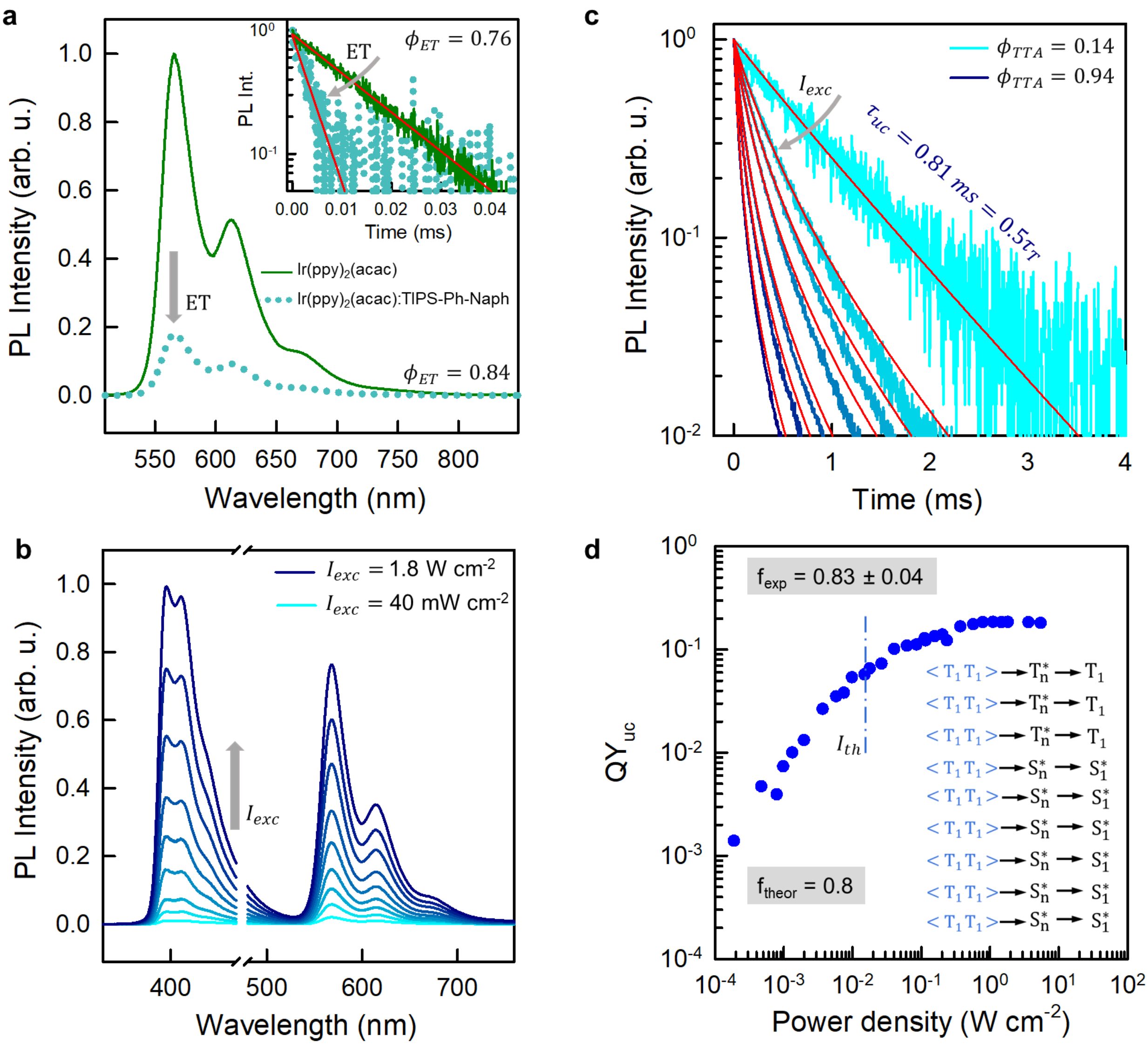

**Fig. 3 | Upconversion properties of Ir(ppy)₂(acac) and TIPS-Ph-Naph solution**. **a)** Photoluminescence (PL) spectra of $Ir(ppy)_2(acac)$ (solid line, $10^{-5}$ M) and of $Ir(ppy)_2(acac)$: TIPS-Ph-Naph (dotted line, $10^{-5}$ M:$10^{-3}$ M) deoxygenated solutions in THF, acquired under a cw laser excitation at 473 nm. The inset reports the PL decay traces at 566 nm recorded under pulsed laser excitation at 405 nm, together with the corresponding single exponential fitting curves (red lines). **b)** PL spectra of the $Ir(ppy)_2(acac)$: TIPS-Ph-Naph solution under cw laser excitation at 473 nm upon increasing excitation intensity $I_{exc}$. For the sake of clarity, the excitation stray light was removed (breaks). **c)** PL decay curves of the upconverted emission at 390 nm under a modulated 473 nm laser excitation at 75 Hz as a function of $I_{exc}$. The solid lines are the fit of data with Eq. 3. **d)** Upconversion efficiency $\phi_{uc}$ measured as a function of $I_{exc}$ of the $Ir(ppy)_2(acac)$: TIPS-Ph-Naph solution (blue dots). The blue dashed-dotted line indicates the excitation intensity threshold $I_{th}$. The inset reports the proposed theoretical schematics of the possible outcomes of the $<T_1T_1>$ relaxation when considering nine annihilating triplet couples, along with the expected and experimental f values.

Supplementary Table S6). This value is slightly below the sensitizer's triplet energy of 2.19 eV (Supplementary Fig. S1), allowing the sensitization through energy transfer of the sTTA-UC mechanism. Importantly, considering the calculations uncertainty and the thermal broadening at room temperature, the quantum mechanical modeling results indicate that the energy of the TIPS-Ph-Naph $<T_1T_1>$ complex $2T_1 = 4.28$ eV is resonant with two high energy singlets, $S_4 = 4.27$ eV and $S_5 = 4.30$ eV, respectively, and one triplet state, $T_{12} = 4.28$ eV (Fig. 2d, Supplementary Table S6). Therefore, fast non-radiative $<T_1T_1>$ relaxation may occur with comparable probability across all these states. These observations suggest that the lateral substitutions place the system in an energetic configuration that favors singlet generation upon TTA. If we consider nine annihilating triplet pairs, six dead pairs statistically form a singlet state $S_n$, which then relaxes to the emissive

$S_1$ level. On the other hand, three dead pairs form a $T_n$ state, which relaxes to $T_1$ partially compensating the triplet loss. This means that twelve out of fifteen dead triplets yield a singlet, with an expected f value of 12/15 = 0.80 (*vide infra*, inset of Fig. 3d). It is interesting to note that when the TIPS groups are replaced by phenyl rings, yielding 1,3,5,7-tetraphenylnaphthalene (Ph-Naph), the resulting substitution pattern produces a $T_1$ state that is too high in energy to enable efficient energy transfer (Supplementary Figs. S3-4).

To validate our design strategy, the TIPS-Ph-Naph molecule was employed as TTA annihilator paired to the $Ir(ppy)_2(acac)$ triplet sensitizer in a deoxygenated THF solution, at the concentrations $10^{-3}$ M and $10^{-5}$ M, respectively (Supplementary Fig. S6a). At this concentration the emitter $\phi_{pl}$ slightly decreases to 0.57±0.04 due to self-absorption (Supplementary Fig. S6b). The $\phi_{ET}$ value was measured by comparing the integrated sensitizer's phosphorescence intensity and lifetime with and without acceptor/emitter, resulting in a $\phi_{ET}$ of 0.80±0.06 (Fig. 3a, Supplementary Fig. S7 and Supplementary section 8). We would like to stress that we used this composition showing an incomplete transfer with the explicit purpose of evaluating the f value by measuring independently and with the maximum accuracy all the parameters in Eq. 2. Figure 3b displays the photoluminescence spectra of the dual dye solution recorded under continuous laser excitation at 473 nm, i.e. in the sensitizer absorption band, measured as a function of the excitation intensity $I_{exc}$. It is easy to observe the upconverted emission from TIPS-Ph-Naph peaked at 390 nm, with an energy gain of about 0.56 eV, as well as the residual phosphorescence from the sensitizer peaked at 566 nm. The occurrence of sTTA-UC is verified by measuring the upconverted emission lifetime at 390 nm as a function of $I_{exc}$. As shown in Fig. 3c, the upconverted emission intensity decay kinetics display the expected dependency on $I_{exc}$, which rules the rate and yield of the TTA process.[30,58] At low powers, the upconverted emission intensity $I_{uc}$ follows a single exponential decay with lifetime $\tau_{uc}$ = 0.81 ms, which is half of the annihilating triplets lifetime $\tau_T = k_T^{-1}$ = 1.62 ms.[29] With this parameter it is possible to fit the decay curves measured at different $I_{exc}$ using[30,58]

$$I_{uc} = \left(\frac{1-\phi_{TTA}}{e^{k_T t}-\phi_{TTA}}\right)^2. \qquad \text{Eq. 3}$$

This allows us to extrapolate the corresponding $\phi_{TTA}$ values, which quickly saturate to unity for $I_{exc}$ > 0.27 W cm$^{-2}$ (Supplementary Table S1). Notably, under continuous and high intensity excitation at 10 W cm$^{-2}$, the system is nicely stable and operative up to 50 minutes (Supplementary Fig. S8).

The final demonstration of sTTA-UC is given by looking at the $\phi_{uc}$ dependency on $I_{exc}$ recorded under steady state excitation at 473 nm reported in Fig. 3d. In agreement with time resolved experiments, the $\phi_{uc}$ increases progressively with $I_{exc}$ until reaching a plateau where $\phi_{TTA}$ = 1 and $\phi_{uc}$ takes its maximum value of 0.19±0.01 (Supplementary section 8). Moreover, from these data we can estimate the excitation intensity threshold $I_{th}$ of the upconverting solution. This parameter marks the excitation intensity at which the TTA rate equals the emitter's triplet spontaneous decay rate, so it is a sample-specific figure of merit that indicates the beginning of the high efficiency upconversion regime. The $I_{th}$ value can be estimated as the excitation intensity at which $\phi_{TTA}$ = 0.5 and so $\phi_{uc}$ is half of its maximum. In our case, the solution shows a threshold of 14.7 mWcm$^{-2}$ ($3.5\times10^{16}$ ph s$^{-1}$ cm$^{-2}$ at 473 nm), in agreement with the predicted value (Supplementary Fig. S9 and Supplementary section 8).[59]

By applying in Eq. 2 all the parameters measured in independent measurements, we estimated the f factor of TIPS-Ph-Naph as large as 0.83±0.04 in excellent agreement with the predicted value 0.80. This is a remarkable result, which suggests that thanks to the optimized electronic structure of TIPS-Ph-Naph almost all the annihilating triplets produce an upconverted singlet, with a $\phi_{uc}$ limited only by the emitter photoluminescence yield. This picture has been validated by comparing TIPS-Ph-Naph with the 1,4-TIPS-Naph annihilator, which is one of the best studied and performing annihilators for Vis-to-UV upconversion.[32,33,55] In this case, the quantum mechanical modeling of 1,4-TIPS-Naph performed points out that the annihilation complex energy is resonant with two triplet states and only weekly resonant with a high energy singlet state (Supplementary Table S7). This suggests that the formation of a triplet by TTA is intrinsically favored over singlet generation, with a theoretical maximum value of 0.5 and an experimentally observed f between 0.32 and 0.54.[32,33] Therefore, despite the better photoluminescence yield, the 1,4-TIPS-Naph ability as TTA upconverter is significantly limited by its intrinsic electronic properties. Comparative transient absorption experiments (Supplementary Figs. S11-12) confirm the more efficient $S_1$-to-$T_1$ ISC in TIPS-Ph-Naph than in 1,4-TIPS-Naph, which is consistent with the lower value observed for $\phi_{pl}$ (0.59 vs. 0.72, respectively), whilst no unexpected interplay between triplets and singlets manifolds is pointed out.

The technological potential of the proposed upconversion system has been tested under broadband non-coherent excitation to evaluate its effectiveness when coupled to an ideal solar-powered device with a bandgap in the blue spectral region, such as a photocatalytic water splitting cell.[60,61] For this experiment, we employed a solution with a higher Ir(ppy)$_2$(acac) concentration of $10^{-4}$ M to maximize the absorption ability and a TIPS-Ph-Naph concentration of $10^{-2}$ M to achieve a complete ET (Supplementary Fig. S10). Fig. 4a shows the emission spectrum of the lamp employed as excitation source and its overlap with the upconverting solution absorption spectrum (Methods). Fig. 4b reports the integrated upconverted emission intensity recorded as a function of $I_{exc}$. The plot displays the expected power dependency of the upconverted emission intensity on $I_{exc}$ in solution.[59,62] At low excitation intensities, we observed the quadratic relationship characteristic of the bimolecular TTA mechanism, which then shifts to a linear trend at higher intensities where $\phi_{TTA}$ = 1 and $\phi_{uc}$ is maximum at 0.19±0.01. Here the $\phi_{uc}$ is only limited by the increased self-absorption of the upconverted emission at the high emitter concentration employed (Supplementary Fig. S6b, $\phi_{pl}$ 0.47±0.04). The intersection of the asymptotic trends marks the sTTA-UC threshold $I_{th}$. Notably, thanks to the long lifetime of the annihilator triplets, we observed an $I_{th}$ of 0.71 mW cm$^{-2}$ (Fig. 4b), which is one order of magnitude lower than the solar irradiance of 8.5 mW cm$^{-2}$ integrated over the sensitizer's absorption band under the AM1.5 condition.[18] This low threshold value obtained under non-coherent broadband excitation, together with the high upconversion yield achieved by means of the successful molecular engineering strategy proposed to maximize the singlet production from TTA, firmly strengthen and expand the technological interest of sTTA-UC systems.

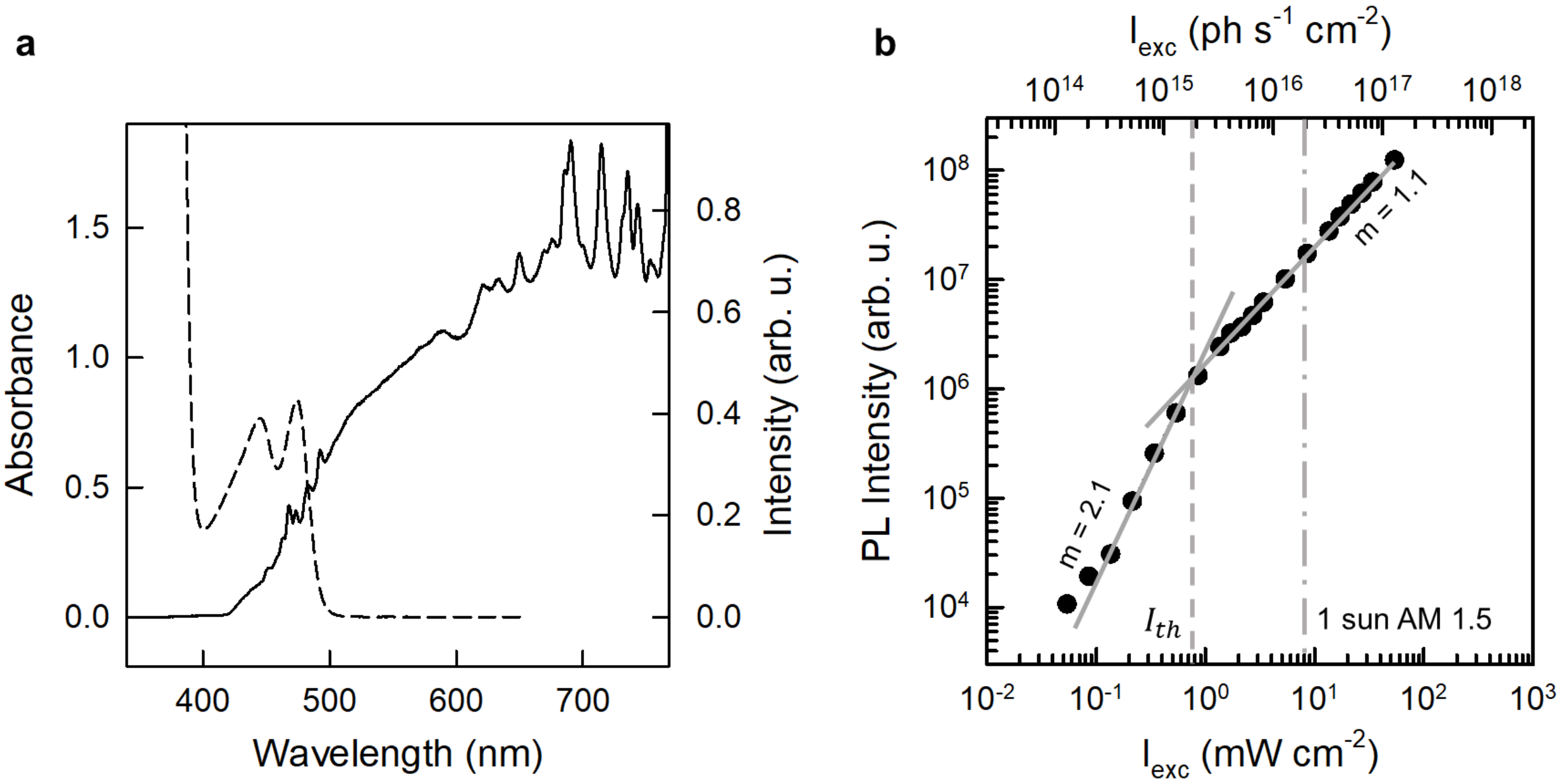

**Fig. 4 | Photon upconversion under broadband excitation**. **a)** Absorption spectrum of Ir(ppy)$_2$(acac): TIPS-Ph-Naph deoxygenated solution in THF ($10^{-4}$ M:$10^{-2}$ M, dashed line) and emission spectrum of the white lamp employed to excite the system. **b)** Integrated upconverted emission intensity recorded as a function of the broadband excitation intensity $I_{exc}$ integrated over the Ir(ppy)$_2$(acac) absorption band between 410 nm and 500 nm. The solid lines are the fitting curves with quadratic and linear dependency on the excitation intensity, respectively, with angular coefficients $m$. The dashed line marks the sTTA-UC excitation threshold intensity $I_{th}$, while the dashed-dotted line marks the solar irradiance integrated over the Ir(ppy)$_2$(acac) absorption band between 410 nm and 500 nm under the AM1.5 condition, corresponding to 8.5 mWcm$^{-2}$.

## DISCUSSION

In summary, efficient Vis-to-UV photon upconversion of broadband non-coherent visible light has been achieved through TTA, by exploiting a newly designed emitter/annihilator paired with a visible-light absorber and triplet sensitizer in solution. The system proposed shows an upconverted emission peaked at 390 nm with a 0.56 eV anti-Stokes shift from the excitation energy. The upconverting solution features a remarkable conversion yield up to 0.19 with an activation excitation threshold which is more than one order of magnitude below the solar irradiance in the AM1.5 condition, matching key technological requirements. The high upconversion yield is achieved by means of a careful engineering of the annihilator excited state energies, which intrinsically promote the formation of high energy singlet states from the triplet-triplet collisional complex formed during TTA. We obtained a remarkable 0.83 yield of singlet formation upon TTA, meaning that almost all the energy stored in the sensitized triplets is effectively exploited. This result along with the spectroscopic studies performed clearly indicate that the modulation of high excited-state energies plays a key role in optimizing the sensitized upconversion performance, pointing out an effective general design strategy, that is in contrast with what pursued so far. The classical approach of designing rigid annihilators with a low density of excited states is meant to avoid energetic singlet/triplet resonances as they statistically limit the generation yield of upconverted singlets through TTA. However, this molecular scenario is rather challenging to accomplish, especially for small-sized systems working in the UV range. Here, we demonstrated that more flexible annihilators with a rich density of high-energy states and tuned resonances work as excellent upconverters affording an almost unit singlet generation yield. This paves the way to a plethora of easier molecular structure modifications, significantly broadening the room for improvement not only in the photon upconversion field, but also for other advanced applications such as OLEDs[63] and scintillation detectors,[64] where breakthrough advances are now required, which can be achieved only by a deeper understating and exploitation of the fundamental electronic properties of luminescent molecular semiconductors.

# Methods

**Materials.** The synthesis of 1,5-bis[(triisopropylsilyl)ethynyl]-3,7-diphenylnaphthalene (TIPS-Ph-Naph) and 1,3,5,7-tetraphenylnaphthalene (Ph-Naph) as well as their structural characterizations, are detailed in the Supplementary Information.

**Material Synthesis.** All reagents were used as received from commercial sources without further purification. Commercial Anhydrous DMF and triethylamine were used. Toluene was distilled over Na/benzophenone. Kept over activated 3 Å molecular sieves. All reactions were carried out under argon. $^{1}$H and $^{13}$C NMR spectra were recorded at room temperature on Brucker Avance-300 MHz NMR spectrometer. $^{1}$H NMR spectra were recorded at 300 MHz and $^{13}$C NMR spectra were recorded at 75 MHz. Chloroform residual peak was taken as internal reference at 7.26 ppm for $^{1}$H NMR and 77 ppm for $^{13}$C NMR. High-resolution mass spectra were obtained by using Waters Xevo Q-Tof using positive mode.

**Quantum Chemical Calculations.** All calculations were performed using Gaussian 16 (Revision C.02) package. The density functional theory (DFT) method was employed for geometry optimization at the B3LYP/6-311+G(d,p) level. To ensure numerical accuracy, a fine integration grid was used. Model systems were employed in which the triisopropylsilyl (TIPS) groups were replaced by trimethylsilyl (TMS) groups. The optimized geometries were confirmed to be local minima by frequency analysis at the same level of theory, with no imaginary frequencies. The time-dependent DFT (TD-DFT) calculations were conducted at the B3LYP/6-311+G(d,p) level for excited-state calculations.

**Spectroscopic studies.** All spectroscopic studies were carried out using THF solutions in quartz cuvettes with 1 mm optical path. The energy transfer and upconversion studies were performed on solutions prepared in a glove box with oxygen level < 1 ppm and there sealed to prevent oxygen contamination. All the absorption spectra were measured with a Agilent Cary 60 spectrometer. The emitters' photoluminescence spectra were acquired using a Varian Cary Eclipse fluorescence spectrometer, where the sample excitation wavelength was selected from the emission of the integrated Xe lamp, while the sample emission was detected by the integrated phototube (PMT). The phosphorescence and upconverted emission spectra were recorded by exiting the samples with a cw 473 nm laser and collecting the signal with a Si charge-coupled device (CCD, Jobin-Yvon Sincerity) interfaced with a Jobin-Yvon Triax 190 monochromator operating with a 300 lines/mm grating.

The measurements under broadband excitation were acquired using a non-coherent broadband white lamp (Xe 900, Edinburgh Instruments) equipped with a 380 nm long pass filter to remove the UV light from the excitation.

The emitters' photoluminescence time decay dynamics were measured using a Ultrafast Systems' tunable APOLLO-Y optical parametric amplifier operating at 340 nm. The emission signal was detected by a PicoQuant PMA Hybrid Series-07, processed through a 74100 Cornerstone 260 ¼ m VIS-NIR monochromator (ORIEL), coupled to a PicoHarp 300 time-correlated single photon counting unit.

The phosphorescence and upconverted emission time decay dynamics were acquired with a PMT (Hamamatsu R5509-73) coupled to time-correlated single-photon counting (TCSPC) electronics (time resolution ~300 ps), exciting the systems with a 405 nm picosecond-pulsed diode laser (Edinburgh EPL405), or with the cw 473 nm laser, modulated via a square wave signal (500 Hz) generated by a TTi TG5011 waveform generator.

Ultrafast transient absorption measurements were performed on Ultrafast Systems' Helios TA spectrometer. The laser source was a 10 W Hyperion amplified laser operated at 2.142 kHz producing ~260 fs pulses at 1030 nm and coupled with an independently tunable APOLLO-Y optical parametric amplifier from the same supplier that produced the excitation pulses at 370 nm (3.35 eV) and 350 nm (3.54 eV). The probe beam was a white light supercontinuum in the UV or VIS spectral range.

For power dependent measurements, the excitation intensity was modulated using neutral density filters. The laser spot size was estimated by means of the knife-edge method, and the laser power incident on the samples was measured with a Thorlabs PM100USB power meter coupled to an S120VC sensor head.

All the photoluminescence spectra reported take are corrected by the detectors spectral response.

**Data availability**
The data supporting the plots and all other findings of this study are available from the corresponding authors upon reasonable request.

**Acknowledgements**
AM and AR acknowledge funding from the National Plan for NRRP Complementary Investments (PNC), project n. PNC0000003 – AdvaNced Technologies for Human-centrEd Medicine (ANTHEM). Additional support to AR was provided by the Italian Ministry of Research through the PRIN2022 program (project Luminance, 2022E42PMA), funded by the European Union - Next Generation EU initiative.

**Author contributions**
M. F. and A. M. conceived the project. M. F. synthesized the new molecules. A.R. performed the upconversion mechanism investigation under A.M. supervision. M.M. performed the quantum mechanical calculations. All authors wrote and revised the paper.

**Competing interests**
The authors declare no conflicts of interest.

**Additional information**
Supplementary information